\documentclass[useAMS,natbib]{mn2e}
\usepackage{amssymb,amsmath}
\usepackage{graphicx}

\title[dbp emitter and x-shaped radio source]
{SDSS J1130+0058 an X-shaped Radio Source With Double-Peaked
Low-Ionization Emission Lines: A binary Black Hole System?}

\author[Zhang et al.]
       {Xue-Guang Zhang$^{1}$\thanks{xguang@astroscu.unam.mx},
        Dultzin-Hacyan D.$^1$,
        Ting-Gui Wang$^2$ \\
       $^1$Instituto de Astronomia, Universidad Nacional Autonoma de
                 Mexico, Apdo Postal 70-264, Mexico D. F. 04510, Mexico \\
       $^2$Center for Astrophysics, Department of astronomy and Applied
                 Physics, University of Science and Technology of China, \\
                 Hefei, Anhui, P.R.China}

\date{}

\pagerange{\pageref{firstpage}--\pageref{lastpage}} \pubyear{2006}
\def\LaTeX{L\kern-.36em\raise.3ex\hbox{a}\kern-.15em
    T\kern-.1667em\lower.7ex\hbox{E}\kern-.125emX}
\begin{document}
\label{firstpage}

\maketitle

\begin{abstract}
    In this paper we study the object SDSS J1130+0058 which  is the only AGN
known to have both double-peaked low-ionization broad emission
lines, and also X-shaped radio structures. Emission from an
accretion disk can reproduce the double-peaked line profile of broad
H$\alpha$, but not the radio structure. Under the accretion disk
model, the period of the inner emission line region is about 230
years. Using a new method to subtract the stellar component from the
data of the SDSS DR4, we obtain an internal reddening factor  which
is less than previously found. The implied smaller amount of dust
disfavors the backflow model for the X-shaped radio structure. The
presence of a Binary Black Hole (BBH) system is the most natural way
to explain ${\it both}$ the optical and radio properties of this
AGN. Under the assumption of the BBH model, we can estimate the BBH
system has a separation of less than 0.04 pc with a period less than 59
years, this may pose some problem to the BLRs sizes, still we conclude
that the BBH model is favored on the basis of the present limited
information.
\end{abstract}


\section{Introduction}

   The detailed structure of the inner parts of Active Galactic Nuclei
(AGN): the black hole, accretion disk and broad emission line
regions (BLRs), cannot be resolved by direct observations, with the
possible exception of indirect evidence from a dusty torus in the
nucleus of NGC4261. The use of reverberation mapping (Peterson 1993;
Netzer \& Peterson 1997; Blandford \& Mckee 1982)
requires the assumption of a rough
geometric structure for the BLRs in order to explain the correlation
between the variations of broad emission lines and those of
continuum.  Moreover, the problem of spatially resolving these
regions, will not be solved in the foreseeable future.  Up until
now, the best approach is the detailed study of broad line profile
shapes and asymmetries from high S/N spectra (Sulentic et al. 2000;
Marziani, Dultzin-Hacyan \& Sulentic 2006).

  There are at least three special types of AGN where the observed
phenomenology has suggested the presence of a binary black hole
system (BBH) in the center: (1) blazars with periodical brightness
variations, (2) AGN with X-shaped radio structures (X-shaped AGN),
and (3) double-peaked broad low-ionization line (dbp) emitters.

  In the prototypical case of the periodically varying blazar OJ 287
(Sillanp\"a\"a et al. 1996a, 1996b; Valtonen et al. 2006a) the
bending of the VLBI (Very Long Baseline Interferometry) jet was
first reported by Vicente et al. (1996). A very small change in
the orientation of the jet is needed to change the Doppler
boosting dramatically, thereby producing long-term periodic
brightness modulations. Sillanp\"a\"a et al. (1988) and  Lehto \&
Valtonen (1996) modeled the periodic outbursts by relating them to
tidally induced mass flows from the accretion disks in a BBH
system with an orbital period of 8.9 years (in the rest frame of
OJ 287). The predictions of the model have been confirmed quite
spectacularly (Valtonen et al. 2006a; Valtonen et al. 2006b).

   The first and most famous dbp emitters are NGC 1097
(Storchi-Bergmann, Nemmen, et al. 2003; Storchi-Bergmann, Eracleous
et al. 1997; Storchi-Bergmann, Eracleous \& Halpern 1995;
Storchi-Bergmann, Baldwin et al. 1993), Arp102B (Chen et al. 1989,
1997; Chen \& Halpern 1989; Halpern et al. 1996;
Antonucci et al. 1996; Sulentic et
al. 1990) and 3C390.3 (Shapovalova et al. 2001; Gilbert et al.
1999). The first natural explanation for the appearance of double
peaked emission lines was also a BBH model (Begelman et al. 1980;  Gaskell
1983). Another model of bipolar outflow  was proposed by Zheng et
al. (1990). The most successful model involved lines originating
from an accretion disk to explain not only the appearance, but
also the observed variations of double-peaked emission lines (Chen
et al. 1989; Chen \& Halpern 1989). The simple accretion disk model has been
modified to better fit the observations to an elliptical disk by
Eracleous et al. (1995), to a warped accretion disk  by Bachev
(1999) and Hartnoll \& Blackman (2000), and to a circular accretion
disk plus spiral arms by Chakrabarti \& Wiita (1994), Hartnoll
\& Blackman (2002) and Karas, Martocchia \& Subr (2001). An
important reason to give up the initial BBH model was the
unreasonable BH masses derived from the variations of the broad
emission lines for three dbp emitters (Eracleous et al. 1997). In
the case of the bipolar outflow model, the observed variations could
not be explained.

   Another special kind of AGN, X-shaped AGN, are classified according to
their extended radio morphology. One of the first explanations for
the distortion of radio structures, particularly for X-shaped and
Z-shaped AGN, was backflow material dominating the bridge regions
(Leahy \& Williams 1984; Worrall et al. 1995). More recently, other
models have also been developed to explain the distortions, such as
the precession of jet axis in a BBH system. It should be noticed
that in the case of OJ287 (the first confirmed periodically varying
blazar), the bending of the VLBI jet was first reported by Vicente
et al. (1996). Objects with bending, misalignment, and wiggling of
extragalactic radio jets (often associated with knots superluminally
moving along different-scale curved trajectory), have been
interpreted in terms of helical structures of the jets. This
structure is likely caused by the precession of the jet in a binary
black hole system (both kinds of models are discussed in e.g.:
Dennett-Thorpe et al. 2002; Merritt \& Ekers 2002;  Liu 2004;
Caproni \& Abraham 2004a, 2004b; Ostorero, Villata \& Raiteri 2004). Up to
the present, there are more than ten X-shaped AGN reported in
the literature.

   It seems only natural to invoke a BBH system to explain the
presence of both dbp lines and X-shaped jets in an AGN. In this
paper, we report the only such object so far known: SDSS
J113021.41+005823 (hereafter, SDSS J1130+0058, found in SDSS DR4
(York et al. 2000; Strauss et al. 2002; Abazajian et al. 2004; Berk
et al. 2001; Scott et al. 2004)). The next section presents the
observational data. In Section 3 we present the discussion and
conclusions. In this paper, the cosmological parameters
$H_{0}=70{\rm km\cdot s}^{-1}{\rm Mpc}^{-1}$, $\Omega_{\Lambda}=0.7$
and $\Omega_{m}=0.3$ have been adopted.

\section{Observed Results}
\subsection{Results at Optical Band}

   SDSS J1130+0058 has been studied by Wang et al. (2003)
with another target name (4C +01.30), as an obscured quasar. Thus,
in this section, we briefly discuss the observed results at optical
band. The apparent optical Petroson magnitudes of SDSS J1130+0058
($z=0.13$) are 18.43, 17.13, 16.19, 15.72, 15.49 at u, g, r, i and z
bands, respectively.  The galactic reddening corrected (with E(B-V)
= 0.022) spectrum of SDSS J1130+0058 with total exposure time of 3600
seconds is shown in Figure 1.

  In order to measure the line parameters more accurately,
the component due to the host galaxy must be removed from the observed
spectrum. We believe that the most convenient way to subtract this
component is the PCA (Principle Component analysis) method
described by Li et al. (2005) and Hao et
al. (2005), using the eigenspectra from pure absorption galaxies from
SDSS or the eigenspectra from stars in STELIB (Le Borgne et al.
2003). Here, we used the method from Hao et al. (2005). The
eigenspectra are calculated by KL (Karhunen-Loeve)
transform for about 926 pure
absorption galaxies selected from SDSS DR2. Then, the first eight
eigenspectra and the spectrum of an A star are used to fit the
absorption properties of the observed spectrum. A power law is used
to fit the continuum from the nuclei. The line spectrum of SDSS
J1130+0058, after the subtraction of the stellar component and power
law continuum, are show in Figure 2. The stellar components and
the power law continuum are also shown in the figure. The best
fitted results for absorption lines, such as the CaII$\lambda3934,
and 3974\AA$ doublet, indicate that the subtraction of stellar
components is reliable. As discussed in Wang et al. (2003), there is
apparent broad H$\alpha$ emission line and weak broad H$\beta$
emission line in line spectrum.

  After the subtraction of the stellar light and continuum, the emission line
parameters are measured. The line parameters were listed in Wang et
al. (2003), however, new eigenspectra and new observed spectrum from
SDSS DR4 are used here, thus, we re-measured the emission lines and
list our new line parameters in Table 1. We fit each narrow emission
line by a narrow gaussian function. For [OIII]$\lambda4959, 5007\AA$
doublet, there are other two gaussian functions for each line,
because of the asymmetry of [OIII]$\lambda4959, 5007\AA$. Due to the
weakness of broad H$\beta$, we use the broad component of H$\alpha$
to fit broad H$\beta$. First, we obtain the broad component of
H$\alpha$ after the subtraction of narrow H$\alpha$,
[NII]$\lambda6548,6583\AA$ and [SII]$\lambda6716,6731\AA$ within the
wavelength range from 6400$\AA$ to 6800$\AA$. In order to get more
accurate narrow components near H$\alpha$, there are two broad
gaussian functions to fit the broad H$\alpha$. Then the narrow
components near H$\alpha$ according to the best fitted results are
subtracted from the observed line spectrum. We have assumed that
broad H$\beta$ and H$\alpha$ have the same line profiles, thus we
can estimate the flux ratio of broad H$\alpha$ to H$\beta$. The best
fitted results for H$\alpha$ are shown in Figure 3. Figure 4 shows
the best fitted results for H$\beta$ by scaling of the observed 
broad H$\alpha$.

  The object SDSS J1130+0058, has been considered as a dbp emitter in
the sample of dbp emitters by Strateva et al. (2003). In our new
sample of dbp emitters (Zhang  et al. in preparation) from SDSS DR4,
this object also is a dbp emitter. For preliminary definition of
criteria see also Zhang, Dultzin-Hacyan \& Wang (2007).

\subsection{Results at Radio Band}

   The radio image at 20cm for SDSS J1130+0058 can be extracted from
Faint Images of the Radio Sky at Twenty centimeter Survey (FIRST,
Becker, White \& Helfand 1995 ), which is shown in Figure 5. The
apparent X-shaped radio morphology can be seen. There are four
regions which strongly emit radio power as listed in Table 2. As
discussed in Wang et al. (2003), the smallest offset of the
photometric center from the four radio regions is about 4.5
arcseconds which is larger than the position uncertainty of SDSS and
FIRST, which indicates that the central component may be a
component from the jet rather than the radio core. The total
integral radio power at 20cm for SDSS J1130+0058 is about
$10^{25.42}{\rm W\cdot Hz^{-1}}$.

   Furthermore, one fundamental parameter of a radio jet, the age
of the jet, is important to understand the physical processes in the
forming  the jet. There are many papers that discuss the age of a
jet, including the dynamical and spectral age. Alexander \& Leahy
(1987) firstly proposed the method to measure the spectral age of a
radio jet (the same methods can be found in Leahy et al. 1989; Liu
et al. 1992; Params et al. 1999). Spectral age depends on the break
frequency $\mu_{br}$, which can be calculated by the spectral index
at different frequencies in various regions. The more accurate
dynamical age is more difficult to calculate and depends on the
model of dynamical expansion (Kaiser \& Alexander 1997; Kaiser 2000;
Kaiser et al. 1997; Parma et al. 1999; Machalski et al. 2006; Tudose
et al, 2006; Kino \& Kawakatu 2005). The limit on the radio data is
not enough to calculate the accurate age. However, we can determine
which one is older. Machalski et al. (2006) have found there is a
strong linear correlation between spectral age and dynamical age
(dynamical age is about two times the spectral age). The spectral
age $t_{s}$ can simply depend on the distance between the lobe edge
and the core $x$: $t_{s}\propto x$, considering the relation between
$x$ and $\mu_{br}$ shown in Parma et al. (1999). Thus as a simple
empirical result, we can say that the double-jet in the N-S direction is older than
the one in E-W direction, because of the longer projected distance
in N-S direction.

\section{Discussions and Conclusions}
\subsection{An obscured AGN?}

 The internal reddening factor can be determined by the value of
the Balmer decrement. The flux ratios of the Balmer emission lines
are about 3.78  and 6.52 for the narrow and broad components
respectively. It should be pointed out, however, that the intrinsic
Balmer decrement is not a definitive value for AGN, because case B
recombination is not necessarily valid (particularly for the BLR).
The value of 3.1 can be expected by Case B recombination. From a
sample of quasars with smaller contribution from the host galaxy, the
mean value is about 3.5 (Greene \& Ho, 2005b). If we assume the
intrinsic Balmer decrement as 3.1, the internal reddening factors
are about E(B-V)=0.64 for the broad Balmer emission lines, and
E(B-V) = 0.17 for the narrow Balmer emission lines. The Balmer
decrement is not as large as the one determined by Wang et al.
(2003).

   In order to verify if the internal reddening factor leads to
luminosities which agree with well established scaling relations,
we have used the tight correlation between continuum luminosity at
5100$\AA$ and the luminosity of H$\alpha$ found by Greene \& Ho
(2005b). The observed luminosity of H$\alpha$ including the narrow
component of H$\alpha$ is about 1.16$\times10^{42}{\rm erg\cdot
s^{-1}}$ ( $9.08\times10^{41}{\rm erg\cdot s^{-1}}$ for broad
component and $2.47\times 10^{41}{\rm erg\cdot s^{-1}}$ for narrow
component). The continuum luminosity at 5100$\AA$ from the nucleus
can be measured from the power law continuum,
$L_{5100\AA}=1.09\times10^{43}{\rm erg\cdot s^{-1}}$. The total
continuum luminosity at 5100$\AA$ from the observed spectrum is
approximately $4.79\times10^{43}{\rm erg\cdot s^{-1}}$. After the
internal reddening correction with $E(B-V) = 0.64$, the intrinsic
continuum luminosity at 5100$\AA$ from the nucleus is about
$8.03\times10^{43}{\rm erg\cdot s^{-1}}$, and the intrinsic
luminosity  L$_{H\alpha}$ is about $4.01\times10^{42} {\rm
erg\cdot s^{-1}}$ (the luminosity of narrow H$\alpha$ has been
corrected by the factor E(B-V)=0.17). The relation between the
corrected values are consistent with the expectation from the
correlation found by Greene \& Ho (2005b):
\begin{equation}
L_{H\alpha}=5.25\times10^{42}(\frac{L_{5100\AA}}{10^{44}{\rm
erg\cdot s^{-1}}})^{1.157} {\rm erg\cdot s^{-1}}
\end{equation}

   This consistency  confirms
that the method of subtracting stellar components is valid and the
calculated internal reddening factor is a reasonable estimate.
SDSS J1130+0058 is indeed an obscured AGN with intermediate internal
reddening. The smaller value for internal reddening found in this
paper (as compared to Wang et al. 2003) implies a smaller amount of
internal dust, and thus a lower density for the internal regions.
However, the backflow model requires a higher gas pressure in
the direction of the galaxy, as shown in other X-shaped radio
sources there are nearly no detectable broad emission lines because
of the higher internal reddening. This does not favor the
backflow model for the X-shaped radio structure in SDSS J1130+0058.

\subsection{BH Masses and Accretion Rate}

   There are mainly two methods to estimated the central
BH masses: One is from the correlation between the masses of the
bulge and the central Black Hole following  the strong correlation
between BH masses and stellar velocity dispersion (Tremaine et al.
2002; Ferrarese \& Merritt 2001; Gebhardt, Bender et al. 2000) (or
line width of narrow emission lines, Greene \& Ho 2005a; Nelson \&
Whittle 1995, 1996; Nelson 2000), and the correlation between
absolute magnitude of the bulge and BH masses (Kormendy \& Gebhardt
2001) etc.  The other one is from the assumption of virialization
(Kaspi et al. 2000; Peterson \& Wandel 1999; Peterson et al. 2004;
Onken et al. 2004; Sulentic et al. 2006), based on the line width of
broad emission lines and the size of broad emission line regions estimated
from continuum luminosity (Kaspi et al. 2000;
2005). As shown in Kazantzidis et al. (2005), the central
masses of merging galaxies should also obey the strong correlation
between BH masses and stellar velocity dispersions, if they undergo
a dissipational process with star formation mergers. Moreover,
according to the hierarchical merging model, we think the BBH system
is common in the center of AGN. Even in the sample used to obtain
the strong relation $M_{BH} - \sigma$ (Tremaine et al. 2002;
Ferrarese \& Merritt 2001; Gebhardt, Bender et al. 2000), we cannot
discard the possibility that the centers of the objects have
BBH system. Thus,
whether there is a BBH system in the center of SDSS J1130+0058 or
not, we estimate the BH masses from stellar velocity dispersions.

  How to measure accurately stellar velocity dispersions is
an open question, because of the known problem about the template
mismatch. A commonly used method is to select spectra of several
kinds of stars (commonly, G and K) as templates, and then broaden
the templates by the same velocity to fit stellar features, leaving
the contributions from different kinds of stars as free parameters
(Rix \& Whit 1992). However, more information about stars included
by the templates should lead to more accurate measurement of stellar
features. According to the above mentioned fitting method, we
selected a new template rather than several spectra of G or K stars
as templates. The method of Principle Component Analysis (PCA)
provides a better way to constrict more favorable information from a
series of spectra of stars into several eigenspectra. Thus, we apply
PCA method for 255 spectra of different kinds of stars in STELIB.
Selecting the first several eigenspectra and a three-order
polynomial function for the background as templates, the value of
the stellar velocity dispersion can be measured by means of the method of
minimum $\chi^2$ fit. Figure 6 shows the fitted results for the stellar
features near MgIb$\lambda5175\AA$ by means of the eigenspectra with
broadening velocity $\sigma=142.6\pm22.6{\rm km\cdot s^{-1}}$. The
error in the stellar velocity dispersion is determined from the different
numbers of required eigenspectra. Here, 4 to 8 eigenspectra are used to
fit the stellar features. The central BH masses of SDSS J1130+0058
can be estimated as $M_{BH}=3.83^{+4.44}_{-2.37}\times10^{7}{\rm
M_{\odot}}$ by (Tremaine et al. 2002; Ferrarese \& Merritt 2001;
Gebhardt, Bender et al. 2000):
\begin{equation}
M_{BH} = 10^{8.13\pm0.06}(\frac{\sigma}{200{\rm km\cdot s^{-1}}})^{4.02\pm0.32} {\rm M_{\odot}}
\end{equation}
Moreover, we should select another parameter to estimate the central
BH masses. The convenient way is to select the line width of narrow
emission lines to trace the stellar velocity dispersion in the
bulge. From the line widths of narrow emission lines listed in Table
1, the minimum value of line width ($\sigma_{line}$) is 103.3${\rm
km\cdot s^{-1}}$ for H$\delta$ and the maximum value is 175.6${\rm
km\cdot s^{-1}}$ for [OII]$\lambda3727\AA$. Thus, we can estimated
the BH masses as $9.47\times10^{6} - 8.01\times10^7 {\rm M_{\odot}}$
by the relation of $M_{BH} - \sigma$, which is consistent with the
value from stellar velocity dispersion and confirms the strong
correlation between stellar velocity dispersion and line width of
narrow emission lines.

  The virial BH masses are based on the empirical relation between the
size of BLRs and continuum luminosity. However, the broad emission lines of
SDSS J1130+0058 have special double-peaked profiles. Thus, either
under the accretion disk model or under the BBH model, the size of
BLRs of the object should not obey the empirical relation $R_{BLRs}
- L_{5100\AA}$.

   For accretion rate, we can determine the dimensionless accretion rate
based on the bolometric luminosity and Eddington luminosity to trace
the accretion rate. The bolometric luminosity for AGN with normal
spectral energy distribution, such as normal big blue bump, can be
determined by the continuum luminosity $L_{bol}\sim9\times
L_{5100\AA}$. The Eddington luminosity can be determined by $L_{Edd}
\sim1.38\times10^{38}M_{BH}/{\rm M_{\odot}erg\cdot s^{-1}}$. Thus,
the dimensionless accretion rate can be determined as $\dot{m}\sim
L_{bol}/L_{Edd}\sim0.06 - 0.36$, which is larger than the critical
accretion rate for ADAF (Mahadevan \& Quataert 1997;
Narayan et al. 1995, 1996; Mahadevan 1997).

\subsection{Accretion Disk Model or BBH model for double-peaked H$\alpha$?}

  First, we check whether the BBH model can be valid for SDSS J1130+0058.
From the best fitted  results shown in Figure 3, double-gaussian
functions can completely reproduce the observed double-peaked broad
H$\alpha$. If the BBH model is valid, the observed radial velocity of
each peak of each broad component of H$\alpha$
represents the rotating velocity in the line of
sight, $\upsilon_1\sim1065{\rm km\cdot s^{-1}}$ and
$\upsilon_2\sim-2354{\rm km\cdot s^{-1}}$. Here, the minus means the
component is moving toward us. Then the mass ratio of the
two central BH masses can be roughly determined by
$M_{BH1}/M_{BH2}\sim\upsilon_2/\upsilon_1\sim2.2$, if the separation
between the central two black holes is much larger than the distance between
BLRs and each corresponding black hole. Here, $M_{BH1}$ represents
the black hole with larger BH mass. However, whether the relation
$M_{BH} - \sigma$ is valid for merging galaxies depends on whether
the merger is dissipative or colisionless (Kazantzidis et al. 2005).
Thus, we determine the BH masses using another method, before
considering the total BH masses from stellar velocity dispersion. We
can estimate the BH masses of central two black holes under the
correlation $M_{BH} - L_{5100\AA}$ found by Peterson et al. (2004),
$M_{BH1}\sim6.5\pm1.1\times10^7{\rm M_{\odot}}$ and
$M_{BH2}\sim3.1\pm0.9\times10^7{\rm M_{\odot}}$, which is consistent
with the mass ratio from the ratio of rotating velocity.
The continuum luminosity of each black hole + BLRs system can be
estimated from the luminosity of each broad component of H$\alpha$ by
equation 1, because of the small contributions from narrow
components of H$\alpha$. Under the
binary black hole system model, there is another parameter of
projected angle of the orbital velocity to determine the structure
of each BLR except the inclination angle of the orbit, thus we
select the correlation between $M_{BH} - L_{5100\AA}$ rather than eq
2. The total BH masses, $9.6\pm2.0\times10^7{\rm M_{\odot}}$, determined
from each broad H$\alpha$ is similar to BH masses
$3.83^{+4.44}_{-2.37}\times10^7{\rm M_{\odot}}$ estimated from
stellar velocity dispersion, which is consistent with the predicted result for
merging galaxies by Kazantzidis et al. (2005).

   Under the assumption of the BBH model, we accept the BH masses estimated
from each broad H$\alpha$ for each black hole. We can roughly
estimate the separation of the central two black holes by:
\begin{subequations}
\begin{align}
\Omega^2 & = \frac{G\times(M_{BH1}+M_{BH2})}{r^3} \\
\upsilon_{1} & = \Omega\times\frac{M_{BH2}}{M_{BH1}+M_{BH2}}\times r\times\sin(i)\sin(\phi)
\end{align}
\end{subequations}
where $r$ is the separation between the central two black holes, $i$ is the
inclination angle of the rotating plane, $\phi$ is the projected
angle of the orbital velocity, $\Omega$ is the angular speed,
$\upsilon_{1}$ is the observed rotating velocity in the line of sight of
the clouds in BLRs including $M_{BH1}$.
From the equations above, we can
determine the upper limit value of the separation $r$: $r<40.8\ {\rm
light-days}$, and the upper limit value of orbital period $P$:
$P<59\ {\rm years}$. However, the size of each BLR can be determined
by (Wang \& Zhang 2003):
\begin{equation}
\begin{split}
\log{R_{BLRs}} = &(0.51\pm0.02)\log(\frac{L_{H\alpha}}{10^{44}{\rm erg\cdot s^{-1}}}) \\
&+(2.16\pm0.02) {\rm light-days}
\end{split}
\end{equation}

The sizes of BLRs are about 45.1 light-days and 79.4 light-days for
each BLR, which turns out larger than the maximum value of
separation. In order to keep the completeness of each BLR, the sizes
of the BLRs should not be larger than the separation of the two
black holes. This poses some problem to the BBH model for SDSS
J1130+0058.  It has to be stressed, however, that the standard way
to determine BLR sizes, may not be applicable for a BBH system.

   Another model, the accretion disk model has been applied to
explain the double-peaked H$\alpha$. Here, we select elliptical
accretion disk model (Eracleous et al. 1995), because of the number
of free parameters in the model which is less than those required by
the circular accretion disk plus spiral arms model. The best fitted
results with $\chi^2\sim0.91$ are shown in Figure 7. The inner
radius is about $713\pm107{\rm R_{G}}$, outer radius is about
$6327\pm939{\rm R_{G}}$, eccentricity is about $0.22\pm0.01$, the
inclination angle of the disk is about $39\pm3\degr$. The slope of
line emissivity is about $\epsilon\propto r^{-1.58\pm0.11}$. The
local broadening velocity is about $539\pm56{\rm km\cdot s^{-1}}$.
According to the eccentricity, the period of the inner emission
regions can be estimated as
$P_{pre}\sim10.4\frac{1+e}{(1-e)^{1.5}}M_{6}r_{3}^{2.5}\sim230{\rm
years}$, where $M_6$ is the central BH masses in unit of $10^6{\rm
M_{\odot}}$, $r_3$ is the radius in unit of $10^3{\rm R_G}$.

   More information, particularly about the variations of double-peaked
broad $H\alpha$ to determine the time scale and other characteristics  of
the variations of the lines is needed to verify whether the BBH
model is valid for SDSS J1130+0058. From the theoretical side, a
detailed model of the combined accretion disk emission from two
accretion disks is yet to be developed.

\subsection{Final Considerations}

    The only X-shaped radio source with apparent double-peaked
broad emission lines, SDSS J1130+0058, is indeed an internally
obscured AGN with standard continuum luminosity of a Seyfert 1
galaxy. The presence of internal dust may provide a convenient background
for the backflow of material to produce X-shaped radio
structures. But in this paper we found a smaller value of the
reddening (and thus of the dust contents) than previously reported.
We are not in the position to quantitatively support nor reject the
backflow model, however a natural alternative is the BBH model.
Objects with X-shaped extragalactic radio jets may be the signature
of these binary systems.

    A BBH system can also account for the observed dbp Balmer
emission lines. Except for the problem with the sizes of the BLRs,
the separation between the two black holes under the assumption of
BBH model, is consistent with the theoretical expectation by Merritt
\& Ekers (2002), who found that the separation of the two black
holes should be about 0.01 pc to 1 pc in a time shorter than 1 Gyr,
if the coalescence rate of binary black holes is comparable to the
galaxy merger rate. Under the BBH model, the separation of central
black holes for SDSS J1130+0058 is about less than 0.04 pc. From the
results of a single observation, we cannot rule out the BBH model.
Moreover, the periods of BBH model and accretion disk model are
very different, one is about 230 years and the other less than 59
years, so that long term observations are necessary to understand SDSS
J1130+0058 better. Furthermore, the BBH model and accretion disk
model will predict different variation patterns of broad H$\alpha$.
The BBH model leads to radial velocity of each peak varying in
period. On the other hand, the accretion disk model cannot lead to
the same result. Long term monitoring should provide more
information to discriminate which model for SDSS J1130+0058 is
favored.

   There is another possibility within the BBH system scenario
to explain the double-peaked broad emission lines and the X-shaped
radio structures. The double-peaked broad emission lines are coming
from the accretion disk of only one of the two central black holes.
There is NO BLR around the other black hole. Otherwise, there would
be three or more components of broad H$\alpha$. The X-shaped
radio-structures are produced by the precession of the jet. The main
effect of the existence of the second black hole is to produce
the precession of the radio jet with respect to the accretion disk.
In this case, the predicted behavior is that the center wavelength
of broad H$\alpha$ should vary in period, besides the variation of
line profiles. Long term observations also can provide information
to distinguish this model from the other two models.

\section*{Acknowledgments}
ZXG gratefully acknowledges the postdoctoral scholarships offered by
la Universidad Nacional Autonoma de Mexico (UNAM). D. D-H
acknowledges support from grant IN100703 from DGAPA, UNAM. This
paper has made use of the data from the SDSS projects. Funding for
the creation and the distribution of the SDSS Archive has been
provided by the Alfred P. Sloan Foundation, the Participating
Institutions, the National Aeronautics and Space Administration, the
National Science Foundation, the U.S. Department of Energy, the
Japanese Monbukagakusho, and the Max Planck Society. The SDSS is
managed by the Astrophysical Research Consortium (ARC) for the
Participating Institutions. The Participating Institutions are The
University of Chicago, Fermilab, the Institute for Advanced Study,
the Japan Participation Group, The Johns Hopkins University, Los
Alamos National Laboratory, the Max-Planck-Institute for Astronomy
(MPIA), the Max-Planck-Institute for Astrophysics (MPA), New Mexico
State University, Princeton University, the United States Naval
Observatory, and the University of Washington.

\begin{table*}
\centering
\begin{minipage}{120mm}
\caption{Line Parameters of SDSS J1130+0058}
\begin{tabular}{lccc}
\hline
Name & Center Wavelength &  $\sigma$  & flux \\
     &   ${\rm \AA}$ & ${\rm km\cdot s^{-2}}$ & ${\rm 10^{-17}erg\cdot s^{-1}\cdot cm^{-2}}$ \\
\hline
${\rm [NeV]\lambda 3426\AA}$ & 3426.7 & 125.3$\pm$4.3 & 118.4$\pm$3.8  \\
${\rm [OII]\lambda 3727\AA}$ & 3728.8 & 175.6$\pm$2.8 & 261.9$\pm$4.1  \\
${\rm [NeIII]\lambda 3869\AA}$ & 3869.9 & 114.2$\pm$2.9 & 117.3$\pm$2.9  \\
${\rm [NeIII]\lambda 3967\AA}$ & 3969.4 & 141.3$\pm$9.8 & 39.8$\pm$2.6\\
H$\delta$ & 4103.2 & 103.3$\pm$10.6 & 24.2$\pm$2.3 \\
H$\gamma$ & 4341.9 & 105.1$\pm$5.9 & 53.4$\pm$2.9  \\
${\rm [OIII]\lambda 4363\AA}$ & 4364.9 & 116.4$\pm$20.9 & 15.8$\pm$2.7 \\
HeII$\lambda 4686{\rm \AA}$ & 4687.6 & 117.9$\pm$6.4 & 54.6$\pm$2.9  \\
H$\beta$ (narrow) & 4863.4 & 105.5$\pm$2.1 & 139.8$\pm$2.5 \\
H$\beta$ (broad)  &  &  & $f(H\alpha_{broad})/6.52$ \\
${\rm [OIII]\lambda 5007\AA}$ (normal) & 5008.9 & 106.6$\pm$0.4 & 1402.6$\pm$11.5  \\
${\rm [OIII]\lambda 5007\AA}$ (broad) & 5008.8 & 282.3$\pm$17.1 & 150.6$\pm$11.1  \\
${\rm [NII]\lambda 6583\AA}$ & 6586.3 & 119.6$\pm$1.8 & 289.7$\pm$4.3 \\
H$\alpha$ (narrow) & 6565.6 & 112.5$\pm$0.8 & 540.8$\pm$3.8  \\
H$\alpha$ (broad) & 6563.0 & 2655.2$\pm$22.1 & 2177.8$\pm$27.9  \\
${\rm [SII]\lambda 6716\AA}$ & 6719.3 & 122.5$\pm$3.3 & 119.7$\pm$2.8 \\
${\rm [SII]\lambda 6731\AA}$ & 6733.8 & 129.6$\pm$4.2 & 100.4$\pm$2.7 \\
\hline
\multicolumn{4}{c}{double-gaussian components for broad H$\alpha$} \\
\hline
          & 6514.6 & 1283.8$\pm$69.5 & 508.3$\pm$79.7 \\
          & 6590.8 & 2027.2$\pm$84.6 & 1528.2$\pm$91.8 \\
\hline
\end{tabular}
\\
$\sigma$ for each line is from the gaussian fitting. \\
Line parameters of broad H$\beta$ are estimated by the broad component
of H$\alpha$. \\
$f(H\alpha)$ represents the flux of broad H$\alpha$.\\
The parameters of broad component of H$\alpha$ are from fitted results
by one broad gaussian function for broad H$\alpha$. \\
The parameters of [NII]$\lambda6583\AA$ are from fitted results by
two gaussian functions for broad H$\alpha$.\\
There are two components of [OIII]$\lambda4959, 5007\AA$ doublet, one normal
with not much different line width as that of other narrow emission lines,
one broad component with broader line width than that of other emission lines.
\end{minipage}
\end{table*}

\begin{table*}
\centering
\begin{minipage}{120mm}
\caption{Radio flux of SDSS J1130+0058}
\begin{tabular}{ccccc}
\hline
Name & ra & dec & int flux at 20cm & offset \\
     &  &  & mJy &  arcminute \\
\hline
FIRST J113021.6+005820  & 11 30 21.6330 & +00 58 20.000  & 152.94 & 0.075 \\
FIRST J113020.9+005828  & 11 30 20.9480 & +00 58 28.210  & 292.24 & 0.144 \\
FIRST J113022.4+005820  & 11 30 22.4490 & +00 58 20.740  & 77.58  & 0.262 \\
FIRST J113020.1+005827  & 11 30 20.1810 & +00 58 27.110  & 43.96  & 0.315 \\
\hline
\end{tabular}
\end{minipage}
\end{table*}

\newpage
\begin{figure}
\includegraphics[width = 84mm]{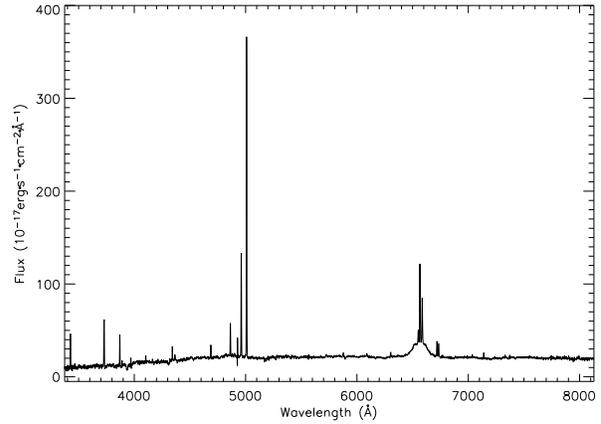}
\caption{The observed spectrum of SDSS J1130+0058.
}
\end{figure}

\begin{figure}
\includegraphics[width = 84mm]{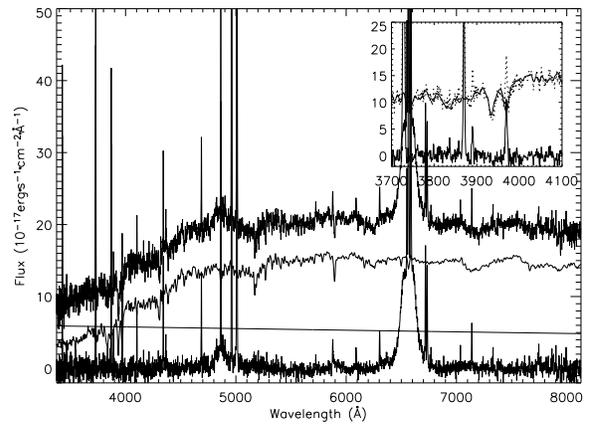}
\caption{
The spectrum of SDSS J1130+0058. The upper one is the observed spectrum.
The middle two are the components of stellar lights and the power
continuum. The bottom one
is the line spectrum after the subtraction of stellar lights and
continuum. The detailed results for CaII$\lambda3934, 3974\AA$ are
shown in top right corner, the dotted line represents the observed
spectrum, solid lines is the best fitted results and line spectrum in
this region.
}
\end{figure}

\begin{figure}
\includegraphics[width = 84mm]{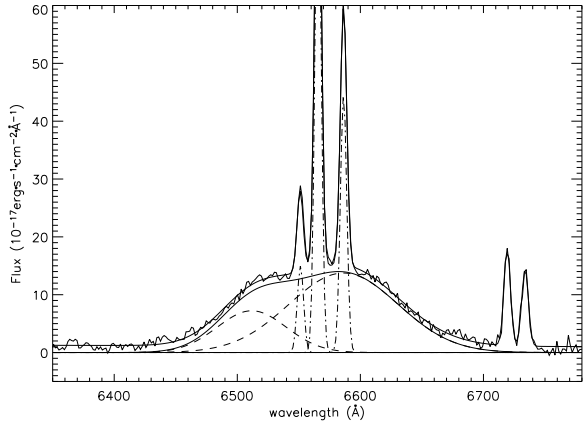}
\caption{The best fitted results for H$\alpha$. Thin solid
line represents the line spectrum. Tick solid line is the
best fitted results. The dashed lines are the two broad components
for broad H$\alpha$. Dot-dashed lines are the narrow emission lines.
The fitted broad H$\alpha$ is shown as solid line under the observed
line spectrum.
}
\end{figure}

\begin{figure}
\includegraphics[width = 84mm]{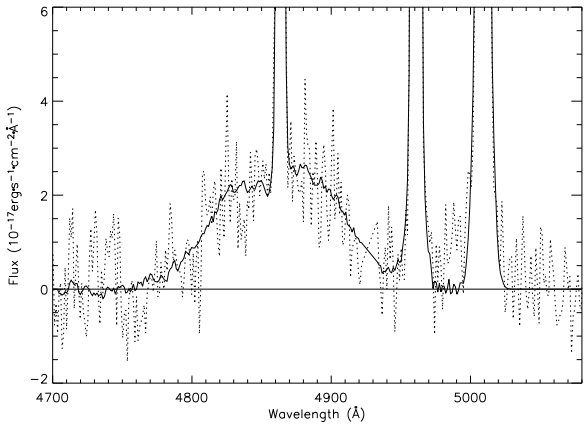}
\caption{The best fitted results for H$\beta$. Dotted
line represents the line spectrum. Solid line is the
best fitted results by scaling of the observed broad H$\alpha$.
}
\end{figure}

\begin{figure}
\includegraphics[width = 84mm]{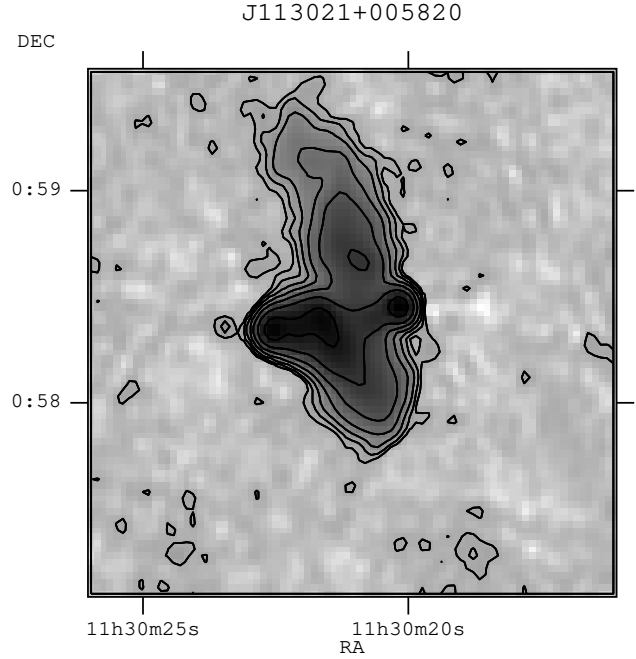}
\caption{The radio image at 20 cm of SDSS J1130+0058.
}
\end{figure}

\begin{figure}
\includegraphics[width = 84mm]{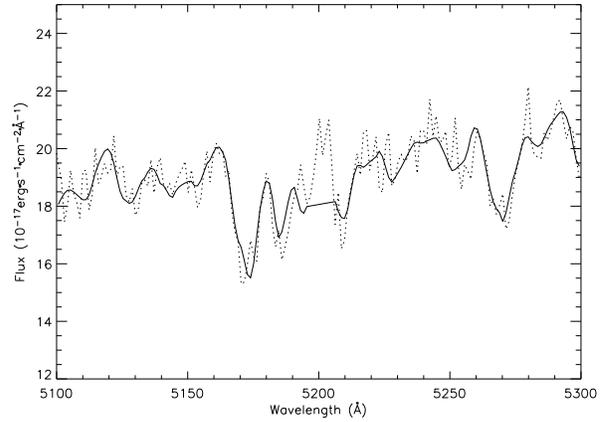}
\caption{ The fitted results for stellar features near
MgIb$\lambda5175\AA$. The dotted line represents the observed
spectrum. The solid line represents the best fitted results with
$\chi^2\sim0.97$ by eight eigenspectra from spectra of stars. }
\end{figure}

\begin{figure}
\includegraphics[width = 84mm]{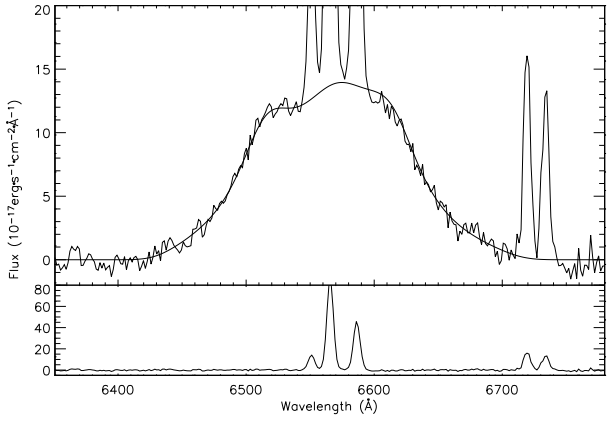}
\caption{
The best fitted results for double-peaked broad H$\alpha$ by
elliptical accretion disk model. The dotted line represents the line
spectrum, solid line is the best fitted results for broad H$\alpha$.
The lower pannel shows the line spectrum after the subtraction of broad
component of H$\alpha$.}
\end{figure}

\end{document}